\documentclass[12pt]{article}
\usepackage{amssymb,amsmath}

\usepackage{curves}
\usepackage{epic}
\usepackage{eepic}
\usepackage{epsfig}

\newcommand{\dd}{\mbox{\rm d}}
\newcommand{\wg}{\wedge}
\newcommand{\gam}{\gamma}

\newcommand{\ddg}{\ddagger}

\newcommand{\DD}{\mbox{\rm D}}

\newcommand{\p}{\partial}

\newcommand{\be}{\begin{equation}}
\newcommand{\bear}{\begin{eqnarray}}

\newcommand{\ear}{\end{eqnarray}}
\newcommand{\ee}{\end{equation}}
\newcommand{\lbl}{\label}
\newcommand{\bi}{\bibitem}
\newcommand{\ci}{\cite}

\newcommand{\vs}{\vspace}

\begin{document}

\begin{center}

\
\vs{1cm}

\baselineskip .7cm

{\bf \Large On the Stueckelberg Like Generalization of General Relativity} \\

\vs{2mm}

\baselineskip .5cm
Matej Pav\v si\v c

Jo\v zef Stefan Institute, Jamova 39, SI-1000, Ljubljana, Slovenia; 

email: matej.pavsic@ijs.si

\vs{3mm}

{\bf Abstract}
\end{center}

\baselineskip .43cm
{\small
We first consider the Klein-Gordon equation in the 6-dimensional space $M_{2,4}$
with signature $+ - - - - +$ and show how it reduces to the Stueckelberg
equation in the 4-dimensional spacetime $M_{1,3}$. A field that satisfies the
Stueckelberg equation depends not only on the four spacetime coordinates
$x^\mu$, but also on an extra parameter $\tau$, the so called evolution time.
In our setup, $\tau$ comes from the extra two dimensions. We point out
that the space $M_{2,4}$ can be identified with a subspace of the 16-dimensional
Clifford space, a manifold whose tangent space at any point is the Clifford algebra
Cl(1,3). Clifford space is the space of oriented  $r$-volumes, $r=0,1,2,3$,
associated with the extended objects living in  $M_{1,3}$.
We consider the Einstein equations that describe a generic curved
space $M_{2,4}$. The metric tensor depends on six coordinates. In the presence
of an isometry given by a suitable Killing vector field, the metric tensor depends
on five coordinates only, which include $\tau$. Following the formalism of
the canonical classical and quantum gravity, we perform the 4 + 1 decomposition
of the 5-dimensional general relativity and arrive, after the quantization,
at a generalized
Wheeler-DeWitt equation for a wave functional that depends on the 4-metric
of spacetime, the matter coordinates, and $\tau$. Such generalized theory
resolves some well known problems of  quantum gravity, including
``the problem of time".}

\baselineskip .55cm

\section{Introduction}
\subsection{The problem of time in quantum gravity}

Despite being a very successful theory at the classical level, general
relativity has turned out to be problematic when attempted to be consistently
quantized. Amongst others, there is the so called `problem of time'
(for a recent review see \,\ci{time}).
This can be seen if we perform the canonical quantization. If we start from
the Einstein-Hilbert action, and perform the $1+3$ Arnowith-Deser-Misner
(ADM) decomposition of spacetime, $M_{1,3} = \mathbb{R} \times \Sigma$,
then the action of general relativity can be cast into the `phase space'
form\,\ci{PhaseSpaceGrav,ADM}
\be
I[q_{ij} ,p^{ij} ,N,N^i ]\, = \int {dt\,d^3 x\,
\,\left[ {p^{ij} \dot q_{ij} \, - \,N\,H(q_{ij} ,p^{ij} ) 
- N_i H^i (q_{ij} ,p^{ij} )} \right]} .
\lbl{1.1}
\ee
Here $q_{ij}$, $i,j=1,2,3$,  is a 3-metric on a space hypersurface
$\Sigma$, and $p^{ij}$ is the corresponding canonically conjugate momentum,
whilst $N$ and $N_i$ are, respectively, laps and shift functions having
the role of Lagrange multipliers leading to the constraints
\be
      {\cal H} (q_{ij} ,p^{ij} )\, \approx 0, ~~~~{\rm and}~~~~~~~ 
      {\cal H}^i (q_{ij} ,p^{ij} )\, \approx 0 ,
\lbl{1.2}
\ee
which are associated with the diffeomorphism invariance of the original
Einstein-Hilbert action. The Hamiltonian is a linear combination of
constraints and the evolution is a pure gauge. There is no physical
evolution time in such a theory.

Upon quantization, the above constraints become the wave functional
equations. For instance, the first constraint becomes the
Wheeler-DeWitt equation
\be
   {\cal H}(q_{ij} , - i{\textstyle{\delta  \over {\delta q_{ij} }}})\,
   \Psi [q_{ij} ]\, = 0
\lbl{1.3}
\ee
whilst the second set of constraints become
\be
    {\cal H}^i  (q_{ij} , - i{\textstyle{\delta  \over {\delta q_{ij} }}})\,
   \Psi [q_{ij} ]\, = 0 .
\lbl{1.4}
\ee
We see that in quantum theory there is no spacetime, but only space $\Sigma$,      ,
because the wave function(al) depends only on 3-geometry, represented by
$q_{ij}$. Thus, in addition to the absence of an external time, we have
also the problem of the disappearance of spacetime.

\subsection{A possible remedy: the Stueckelberg theory}

In the Stueckelberg theory\,\ci{Stueckelberg}, besides the
four spacetime coordinates  $x^\mu$,
there is an extra parameter $\tau$. The coordinate $x^0 \, \equiv \,t$
is not the `evolution parameter'. The evolution parameter is $\tau$,
considered to be a universal ``world time".

In quantum theory of a `point particle', the wave function is
\be
    \psi(\tau, x^\mu),
\lbl{1.5}
\ee
and is normalized according to $\int \dd^4 x \,\psi^* \psi =1$.
We will show how $\tau$ arises from extra two dimensions, one space like
and one time like dimension.

Then we will show
that `extra dimensions' need not be the `true' extra dimensions, i.e., some
extra dimensions in addition to four spacetime dimensions, but can be
associated with the space of matter configurations. A particular case
of such configuration space is Clifford space, a manifold whose tangent
space at any point is the Clifford algebra
Cl(1,3). Clifford space is the space of oriented  $r$-volumes, $r=0,1,2,3$,
associated with the extended objects living in  $M_{1,3}$. In this paper
we focus our attention to a 6-dimensional subspace, $M_{2,4}$,
of Clifford space. We consider the Einstein equations that describe
a generic curved space $M_{2,4}$. Then we perform the ADM-like 1+4 decomposition of
a 5-dimensional subspace $M_{2,3}$ of $M_{2,4}$, our argument being
that the additional dimension can be neglected in the presence
of an isometry given by a suitable Killing vector field, because then
the metric tensor depends on five coordinates only.

We will show how in the quantized theory the problems of time and of spacetime
disappear in such a generalized theory. The latter problem does not occur,
because the wave functional now depends on spacetime 4-geometry,
represented by the metric $g_{\mu \nu}$. The problem of time we resolve
by adding a suitable matter part to the action.

\section{Klein-Gordon equation in 6D}

Let us consider the action for the massless Klein-Gordon field in 6-dimensions:
\be
 I[\phi,\phi^*] 
 = \int \dd^6 x\,\,\partial _M \phi ^* \,\partial ^M \phi 
\lbl{2.1}
\ee
where $\phi=\phi(x^M),~M=0,1,2,3,5,6$. Let us split the index $M$ into
a 4-dimensional part and the part due to the extra two dimensions
according to  $M\, = \,(\mu ,\bar M)$, $\mu=0,1,2,3$,
${ \bar M} = 5,6$, and let us assume that the metric has the
following form:
\be
 G_{MN}  =  \begin{pmatrix}
        g_{\mu \nu }&  0 & 0 \\ 
        0 & 0 & -1 \\ 
        0 & -1 & 0 \\ 
 \end{pmatrix}.
\lbl{2.2}
\ee
The latter metric can be transformed into
\be
G\,'_{MN}  =  \begin{pmatrix}
        g_{\mu \nu }&  0 & 0 \\ 
        0 & - 1 &0 \\ 
        0 & 0 & 1 \\ 
 \end{pmatrix} ,
\lbl{2.3}
\ee
which is the pseudo euclidean metric with signature $(+----+)$. 
By inserting the metric (\ref{2.2}) into the action (\ref{2.1}), we
obtain
\be
I[\phi,\phi^*] = \int \dd^6 x\,(g^{\mu \nu } \,\partial _\mu  
\phi ^* \partial _\nu  \phi \,\, 
- \,\,\partial _\tau  \phi ^* \partial _\lambda  \phi \,\,
 - \,\,\,\partial _\lambda  \phi ^* \partial _\tau  \phi) ,
\lbl{2.4}
\ee 
where we have denoted $x^5 \equiv \tau,~x^6 \equiv \lambda$.

Taking the ansatz
\be
\phi  = e^{i\Lambda \lambda } \psi (\tau ,x^\mu  ) ,
\lbl{2.5}
\ee
where $\Lambda$ is a constant, we have
\be
I[\psi,\psi^*] = \int \dd \tau \, \dd^4 x \,
\,\,\left [ \partial _\mu  \psi ^* \partial ^\mu  \psi \,\, 
+ \,\,i\Lambda \,(\psi ^* \partial _\tau  \psi \,\, 
- \,\partial _\tau  \psi ^* \psi ) \right ] ,
\lbl{2.5a}
\ee
which is the well known Stueckelberg action.
We have omitted the integration over $\lambda$, because it gives
a constant factor which can be absorbed into the definition of the fields.

Considering the corresponding equations of motion, we find that from
a massless Klein-Gordon equation in 6D
\be
      \partial ^M \partial _M \phi  = 0
\lbl{2.6}
\ee
we obtain the Stueckelberg equation
\be
i\,\partial _\tau  \psi \,\, = \,\,\,\,\frac{1}{{2\Lambda }}\,
\partial ^\mu  \partial _\mu  \psi 
\lbl{2.7}
\ee
The constant $\Lambda$ comes from the 6th dimension $x^6  \equiv \,\lambda$.
More precisely, $\Lambda$ is an eigenvalue of the canonical momentum
conjugate to $\lambda$. By ansatz (\ref{2.5}), coordinate $\lambda$ is
eliminated from the action, whilst the eigenvalue $\Lambda$ remains.

To sum up, if the signature of two extra dimensions is $(- +)$, and if, instead
in the coordinates $x'^5, x'^6$ in which the metric is diagonal
(Eq.\,(\ref{2.3})),  we work
in the coordinates $x^5 \equiv \tau = \frac{1}{2}(x'^5 + x'^6)$,
$x^6 \equiv \lambda = \frac{1}{2}(x'^5 - x'^6)$ in which the metric is
non diagonal (Eq.\,(\ref{2.2})), then we obtain the Stueckelberg equation for
a wave function that depends on $\tau$  and $x^\mu$. The coordinates $\tau$,
$\lambda$ are analogous to the light cone coordinates $(x^0 + x^1)/2$,
$(x^0 - x^1)/2$.
Notice that, because $\tau$ is like  a `light cone' coordinate,
we have the first derivative of $\psi$  with respect to  $\tau$. 

\subsection{More formal considerations:
Point particle in 6D and its quantization}

Let us consider a classical action for a point particle in 6-dimensional
space:
\be
I[X^M] = \,M_p \int d \sigma (\dot X^M \dot X_M )^{1/2} ,
\lbl{2.8}
\ee
where $M=0,1,2,3,5,6$, and $M_p$ is the particle's mass in 6D.
Here $\sigma$ is a parameter, denoting a point
on the worldline, and $\dot X^\mu = \dd X^\mu/\dd \sigma$.

An equivalent action is a functional of the coordinates  $X^M$, 
the canonically conjugate momenta $P_M$, and a Lagrange multiplier
$\alpha$:
\be
   I[X^M ,P_M,\alpha]\,\, = \,\,\int d \sigma \left( {P_M \dot X^M \, 
- \,\frac{\alpha }{2}(P_M P^M \, - \,\,M_p^2 )} \right) .
\lbl{2.9}
\ee

Varying the latter action with respect to $P_M$, we obtain the
relation between velocities and momenta, $\dot X^M \, = \alpha P^M$.

If we split the coordinates according to
\be
   X^M \, = \,(X^\mu  ,X^{\bar M} )~,~~~~~{\bar M} = 5,6 ,
\lbl{2.10}
\ee
and express four momenta in terms of velocities, $P^\mu={\dot X}^\mu/\alpha$,
then the action (\ref{2.9}) becomes
 \be
  I[X^\mu]\,\, = \,\,\int d \sigma \left( \frac{\dot X^\mu \dot X_\mu}{2 \alpha}
   +{P_{\bar M} \dot X^{\bar M} \, 
- \,\frac{\alpha }{2}(P_{\bar M} P^{\bar M} \, - \,\,M_p^2 )} \right) .
\lbl{2.11}
\ee

The second term in the latter action can be omitted, because by partial
integration it can be transformed into the form
\be
        \int \dd \sigma  \left (  
        \frac{\dd} {\dd \sigma} ( P_{\bar M} X^{\bar M}) 
        -{\dot P}_{\bar M} X^{\bar M}  \right ),
\lbl{2.11a}
\ee
and if we use the equations of motion, $\dot P_{\bar M} =0$, then only the
total derivative term remains.

The third term in eq.\,(\ref{2.11})
can be rewritten in terms of the 4D mass, $m$. Namely, by varying
(\ref{2.9}) we obtain the mass shell constrain in 6D:
\be
 \delta \alpha :\,\,\,\,G^{MN} P_M P_N \, - \,M_p^2 \, = \,0 ,
\lbl {2.11b}
\ee
which can be decomposed according to
\be
  g^{\mu \nu } P_\mu  P_\nu  \, 
+ \,G^{\bar M\bar N} P_{\bar M} P_{\bar N} \, - \,M_p^2 \, = 0 .
\lbl{2.11c}
\ee
From $M_p^2 = P^M P_M = P^\mu P_\mu + P_{\bar M} P^{\bar M}$, we have
\be
    m^2 \, = \,M_p^2 \, - \,P_{\bar M} P^{\bar M} ,
\lbl{2.12}
\ee
where $m^2 \equiv P^\mu P_\mu$. If 6D mass $M_p$ is equal to zero,
then the 4D mass is due to the 5th and the 6th component of momentum only:
\be
    m^2 = - G^{\bar M\bar N} P_{\bar M} P_{\bar N}= 2P_5 P_6 .
\lbl{2.12a}
\ee

So, from eq.\,(\ref{2.11}), using (\ref{2.11a}), and (\ref{2.12}), we obtain the
well known Howe-Tucker action for a massive particle in 4-dimensional
spacetime:
 \be
  I[X^\mu]\,\, = \mbox{$\frac{1}{2}$}
  \int d \sigma \left( \frac{\dot X^\mu \dot X_\mu}{\alpha}
 + \alpha m^2 \right) .
\lbl{2.13}
\ee

Upon quantization, 
the classical constraint (\ref{2.11b}) becomes
the Klein-Gordon equation
\be
     (G^{MN} \hat P_M \hat P_N \, - \,M_p^2 )\,\phi \,\, = 0 ,
\lbl{2.14}
\ee
where $\hat P_M = -i \hbar \p/\p X^M$ is the momentum operators.
We will use unit in which $\hbar =c =1$, and write $\p_M \equiv \p/\p X^M$.

We can decompose eq.\,(\ref{2.14}) into  a 4D and a 2D part:
\be
    ( - \,g^{\mu \nu } \partial _\mu  \partial _\nu  \, 
- \,\,G^{\bar M\bar N} \partial _{\bar M} \partial _{\bar N} \,
 - \,M_p^2 )\phi \, = \,\,0,
\lbl{2.15}
\ee
which gives
\be
    ( - \,g^{\mu \nu } \partial _\mu  \partial _\nu  \, 
 + \,2\,\partial _5 \partial _6 \, - \,M_p^2 )\phi \, = \,\,0 .
\lbl{2.16}
\ee
By the ansatz
\be
   \phi  = \,{\rm{e}}^{iP_6 x^6 } \psi (x^5 ,x^\mu  )\,
\lbl{2.17}
\ee
and by denoting $x^5 \, \equiv \,\tau,~~P_6 \, \equiv \Lambda$,
eq.\,(\ref{2.16}) gives

\be
i\,\frac{\partial }{{\partial \tau }}\,\psi \, 
= \,\frac{1}{{2\Lambda }}\,\left( {g^{\mu \nu } \partial _\mu  \partial _\nu  \, 
+ \,M_p^2 } \right)\psi 
\lbl{2.18}
\ee
If, in particular, the 6D mass  $M_p$  is zero, then
we have the usual Stueclkelberg equation (\ref{2.7}).
Alternatively, $M_p^2$ in Eq.\,(\ref{2.18}) can be eliminated
by the phase change $\psi \rightarrow 
{\rm exp} [-i \, \frac{M_p^2}{2 \Lambda}\, \tau]\, \psi$. 

We have seen that the Stueckelberg equation in which the wave function
depends not only on four spacetime coordinates $x^\mu$, but also on an extra
parameter  $\tau$ (evolution parameter), is embedded in the 6D theory
with one time-like and one space-like extra dimension:
\be
\begin{tabular}[t]{|l|}
\hline
6D space $M_{2,4}$ \\
signature $+----+$\\
\hline
\end{tabular}
\begin{tabular}[t]{l}
\\
$~~~\longrightarrow~~~$\\
\end{tabular}
\begin{tabular}[t]{|l|}
\hline
Stueckelberg theory in $M_{1,3}$ \\
with invariant evolution parameter $\tau$\\
\hline
\end{tabular}
\nonumber
\ee

\vs{1mm}

At this point it is interesting to observe, that an extra time like and
an extra space like dimension are necessary 
in the ``two time" physics\,\ci{Bars}, based on the extended phase space
action that is invariant under local Sp(2) transformations. A special case
of the latter action is the phase space action for a relativistic
point particle in 4-dimensions. Since our phase space action (\ref{2.9})
is in six, and not in four dimensions, this means that in the considered
6-dimensional theory we do not not impose the Sp(2) constraints on
$X^M$ and $P^M$. We can envisage that such constraints are imposed
in a space of a higher dimensionality, and that a particular, gauge fixed,
case of the
higher dimensional, Sp(2) invariant, action, is the phase space action
(\ref{2.9}). Thus, our approach differes from that in refs.\,\ci{Bars}.
We do not impose the Sp(2) constraints in $M_{2,4}$, but we admit
the possibility that such constraints hold in a higher dimensional space
within the context of a theory that is a modification of the two time
physics \,\ci{Bars}.
In such a way it is possible to embed the Stueckelberg theory into
the theory based on the local Sp(2) invariance. 

What about ghosts? It is usually taken for
granted that time like dimensions imply ghosts. But there is another,
not so well known, possibility that is based on an alternative definition
of vacuum \ci{Jackiw}, in which case no ghosts are associated with
time like dimensions. How this works within the context of string
theory and quantum field theory, and how this can resolve the
cosmological constant problem, was shown in Refs.\,\ci{PavsicSaasFee,
PavsicPseudoHarm,PavsicBook}.

A question arises as to what is a physical meaning
of the extra dimensions. This will be discuss in next section.

\section{The space $M_{2,4}$ as a subspace of Clifford space}

Clifford space, $C$, is the space of oriented $r$-volumes, 
$r=0,1,2,3$, associated with extended objects, such as strings/branes
living in spacetime $M_{1,3}$.
 The concept of Clifford space---a manifold whose
tangent space at any point is the Clifford algebra $Cl(1,3)$---has been
discussed in refs.\,\ci{CastroHint}--\ci{PavsicMaxwellBrane}. It was found
that a curved $C$, since being a higher dimensional space, enables the
unifications of interactions \`a la Kaluza-Klein without introducing the
extra dimensions of spacetime. The `extra dimensions' of $C$ are due to the
fundamental extended nature of physical objects, they are the dimensions
of a configurations space. In principle, those degrees of freedom
are not hidden from our direct observation, therefore we do not need to
compactify such `internal' space.
Here we will exploit the fact that the space $M_{2,4}$, used in previous
section, can be identified with a subspace of $C$.

\subsection{Clifford space: a quenched configuration space of extended
objects--branes}

Strings and branes have infinitely many degrees of freedom.
But at first approximation we can consider just  the center of mass,
$x^\mu,~\mu=0,1,2,3$.
Next approximation is in considering the holographic coordinates, $x^{\mu \nu}$,
of the oriented area  enclosed by the string.
We may go even further and search for eventual thickness of the object.
If the string has finite thickness, i.e., if actually it is not a string,
but a 2-brane,
then there exist the corresponding  volume degrees of freedom,
$x^{\mu \nu \rho}$.

In general, for an extended object in $M_{1,3}$, we have 16
coordinates\,\ci{PavsicArena,PavsicMaxwellBrane}
\be
x^M \, \equiv \,x^{\mu _1 ...\mu _r } \,,\,\,\,\,\,\,\,r = 0,1,2,3,4 .
\lbl{3.1}
\ee
They are the projections of $r$-dimensional volumes (areas) onto the
coordinate planes. Although branes have infinitely many degrees of
freedom, we can sample them by a finite set of coordinates $x^M$ that
denote position in a 16-dimensional space. Let us first assume that the
latter space is flat. Then the position can be described by a vector
$x = x^M \gam_M$, where $x^M$ are components, and $\gam_M$ basis vectors.
For the basis vectors we will take the basis elements of the Clifford
algebra $Cl(1,3)$, thus $\gam_M \equiv \gam_{\mu_1} \wg \gam_{\mu_2} \wg...
\wg \gam_{\mu_r}$, $r=0,1,2,3,4$. The vector $x \in Cl(1,3)$, picturesquely
called `polyvector', is an aggregate of $r$-vectors, i.e., of scalars,
vectors, bivectors, threevectors (pseudovectors) and fourvectors
(pseudo scalars). We can now assume that the Clifford algebra $Cl(1,3)$ is a
tangent space of a 16-dimensional manifold, called Clifford space $C$. If
the manifold $C$ is flat, then it is isomorphic to the Clifford algebra
$Cl(1,3)$, which is the tangent space at a chosen point $P\in C$, say
the ``origin". In general, $C$ can have non vanishing curvature,
in which case it is not isomorphic to $Cl(1,3)$.

Coordinates $x^M$ of Clifford space $C$ can be used to model extended objects,
whatever they are. The latter coordinates, the so called
`polyvector coordinates', are a generalization of the concept of center of
mass\,\ci{PavsicArena}. Instead of describing  extended objects in 
``full detail", we
can describe them in terms of the center of mass, area and
volume coordinates. Namely,
a configuration of an extended object, such as a brane, has infinitely
many degrees of fredom, and the space ${\cal M}$ of all possible brane
configurations\,\ci{PavsicBook} is infinite dimensional. A full description
of a brane corresponds to a point in ${\cal M}$ that requires infinitely
many ``coordinates'', i.e., the brane embedding functions $X^\mu (\xi^a)$.
A ``quenched" description of a brane corresponds to a point in $C$ that
needs sixteen coordinates only, i.e., the coordinates $x^M$.
Therefore, the Clifford space, $C$, is a quenched
configuration space for extended objects\,\ci{AuriliaFuzzy}.

Instead of the usual relativity, formulated in spacetime, in which the
interval is
\be
{\rm{d}}s^2 \, =  g_{\mu \nu \,} {\rm{d}}x^\mu  {\rm{d}}x^\nu  
\lbl{3.2}
\ee
let us consider the theory in which the interval is extended to
Clifford space:
\be
{\rm{d}}S^2 \, = \,G_{MN} \,{\rm{d}}x^M {\rm{d}}x^N 
\lbl{3.3}
\ee
where 
${\rm{d}}x^M \, \equiv \,{\rm{d}}x^{\mu _1 ...\mu _r }$, $~~r = 0,1,2,3,4$.

In particular, extended objects
can be fundamental branes.

The line element (\ref{3.3}) can be written as the scalar product
of the Clifford number
\be
{\rm{d}}X\, = \,{\rm{d}}x^M \gamma _M  \equiv 
\,{\rm{d}}x^{\mu _1 \mu _2 ...\mu _r } \gamma _{\mu _1 \mu _2 ...\mu _r \,} 
,\,\,\,\,\,\,\,\,\,\,\,r = 0,1,2,3,4\,
\lbl{3.4}
\ee
with its reverse $\dd X^\ddg$:
\be
{\rm{d }}S^2  \equiv |{\rm{d}}X|^2  
\equiv {\rm{d}}X^\ddag  *{\rm{d}}X = {\rm{d}}x^M {\rm{d}}x^N G_{MN}  
\equiv {\rm{d}}x^M {\rm{d}}x_M .
\lbl{3.5}
\ee
The metric is given by the scalar product of the basis Clifford numbers:
\be
G_{MN}  = \gamma _M^\ddag * \gamma _N \, 
\equiv \,\langle \gamma _M^\ddag  \,\gamma _N \rangle _0 .
\lbl{3.6}
\ee
Reversion, denoted by $\ddg$, is an operation that reverses the order
of vectors in a Clifford product: 
$(\gamma _{\mu _1 } \gamma _{\mu _2 } ...\gamma _{\mu _r } )^\ddag
 = \gamma _{\mu _r } ...\gamma _{\mu _2 } \gamma _{\mu _1 }$. In flat
 Clifford space, $\gamma _M \equiv \gamma _{\mu _1 \mu _2 ...\mu _r \,}$
is the wedge product of basis vectors,
$ \gam_M =\gam_{\mu_1} \wg \gam_{\mu_2} \wg ... \wg \gam_{\mu_r}$, at
every point of $C$. This is not the case in curved $C$.

With the definition (\ref{3.6}) of the metric, signature of $C$ is
$(8,8)$. Therefore, $M_{2,4}$ is a subspace of $C$.

\subsection{Dynamics}

The following action generalizes the action for a point particle
of the ordinary special relativity:
\be
I = M_p \int_{}^{} {d\sigma \,(\eta _{MN} \dot X^M \dot X^N } )^{1/2} ,
\lbl{3.7}
\ee
where $\sigma$ is an arbitrary continuous parameter.
From the latter action we obtain the following equations of motion:
\be
\ddot X^M \, \equiv \,\,\frac{{\,{\rm{d}}^{\rm{2}}
 X^M }}{{{\rm{d}}\sigma ^2 }}\,\, = \,\,0
\lbl{3.8}
\ee
Here $\eta_{MN}$ is the analogue of Minkowski metric with signature $(8,8)$.

Since $X^M$ are interpreted as $r$-volume coordinates, the
equations of motion (\ref{3.8}) imply that the volume (in particular the area)
changes linearly with $\sigma$. If the coordinates $X^M$ sample a brane,
then the above dynamics can only hold for a tensionless brane. For a brane
with tension one has to introduce curved Clifford space and generalize
eqs.\,(\ref{3.7}),(\ref{3.8}) to arbitrary metric with non vanishing
curvature\,\ci{PavsicKaluza,PavsicKaluzaLong,PavsicInsight}.

A worldline $X^M (\sigma)$ in $C$  represents
the evolution of  a  `thick particle'
in spacetime  $M_{1,3}$. In $C$ we have a line, a worldline $X^M (\sigma)$,
whilst in spacetime  $M_{1,3}$, we have a thick line whose centroid line is
$X^\mu (\sigma)$. It describes a thick particle, i.e., an extended object,
in spacetime. The thick particle can be an aggregate of $p$-branes for
various $p=0,1,2,...$\,. But such interpretation is not
obligatory. A thick particle may be a conglomerate of whatever
extended objects that can be sampled by `polyvector' coordinates
$X^M \equiv X^{\mu_1 \mu_2 ... \mu_r}$.

\section{Einstein's equations in $M_{2,4}$}

Let $x^M$, $M=0,1,2,3,5,6$ be coordinates, and
$G_{MN} \, = \,G_{MN} \,(x^M )$ a metric tensor in $M_{2,4}$.
The Einstein-Hilbert action in the presence of a point like source\footnote{
In our interpretation of the space $M_{2,4}$ as a subspace of 
Clifford space $C$, which is a configuration space associated with an
extended object, a point like source in $M_{2,4}$ is a thick source in
4D spacetime $M_{1,3}$.}
reads
\be
\,\,I[X^M ,G_{MN} ] = M_p\int {\rm{d}} \sigma \,(\dot X^M \dot X^N G_{MN} )^{1/2}  
+ \,\,\frac{1}{{16\pi \cal G }}\int {{\rm{d}}^6 x\,} \sqrt{-G} R^{(6)} \,\,
\lbl{4.1}
\ee
If we vary the latter action with respect to $X^M (\sigma)$, we obtain the
geodesic equation,
\be
\,\,\frac{1}{{\sqrt {\dot X^2 } }}\,\frac{{\rm{d}}}{{{\rm{d}}\sigma }}
\left( {\frac{{\dot X^M }}{{\sqrt {\dot X^2 } }}} \right) 
+ \,\,\Gamma _{JK}^M \frac{{\dot X^J \dot X^K }}{{\dot X^2 }} = 0 ,
\lbl{4.2}
\ee
and if we vary it with respect to $G_{MN} (x^M)$, we obtain the
Einstein equations,
\be
  R^{MN}  - \frac{1}{2}G^{MN} R = 8\pi \,{\cal G}
  \int \dd \sigma \, \delta^6 (x-X(\sigma)) \dot X^M \dot X^N
\lbl{4.3}
\ee
We can use eqs.\,(\ref{4.1}),(\ref{4.3}) as an
approximation to a physical situation in which instead of
the $\delta$-distribution we have a distribution due to an extended source.

The 6D Ricci scalar can be written as
\be
   R^{(6)} \, = \,R^{(4)} \, 
      + \,\,{\rm{extrinsic}}\,\,{\rm{curvature}}\,\,{\rm{term}}_{\,5,6} ,
\lbl{4.4}
\ee
where the subscripts 5,6 mean that the extrinsic curvature is due to the
presence of the 5th and 6th dimension.
Instead of performing such  ADM-like $2+4$ decomposition, we will follow
an easier procedure. We will consider a $1+5$ decomposition  in which case
we have
\be
 R^{(6)} \, = \,R^{(5)} \, 
      + \,\,{\rm{extrinsic}}\,\,{\rm{curvature}}\,\,{\rm{term}}_{\,6}
\lbl{4.5}
\ee
If there exist suitable isometries in the 6D space $M_{2,4}$,
and if we choose a suitable 5D subspace $M_{2,3}$, then the extrinsic
curvature terms in eq.\,(\ref{4.5}) can vanish. Namely, the extrinsic
curvature term tells how the hypersurface is bended with respect to the
emebdding space, and it can be bended so that the extrinsic curvature
is zero\footnote{In flat embedding space this means that the hypersurface is
not bended at all.}. 
We will assume that this is the case.
 
The 5D Ricci scalar, in turn, can also be decomposed in an analogous way:
\be
 R^{(5)} \, = \,R^{(4)} \, 
      + \,\,{\rm{extrinsic}}\,\,{\rm{curvature}}\,\,{\rm{term}}_{\,5}
\lbl{4.6}
\ee

In particular, let us consider the ADM-like 1+4 decomposition
$M_{2,3}= \mathbb{R} \times M_{1,3}$, where $M_{1,3}$ is spacetime.
Then the 5D metric can be decomposed as
\be
G_{MN} \, = \,\left( \begin{array}{l}
   {\cal N}_\mu  {\cal N}^\mu +{\cal N}^2  \,,\,\,\,\,{\cal N}_\mu \\ 
 \,\,\,\,\,\,\,\,\,\,{\cal N}_\nu  \,\,,\,\,\,\,\,\,\,\,\,\,\,\,\,\,\,g_{\mu \nu }  \\ 
 \end{array} \right) ~,
 ~~~~  \begin{array}{l}
 M,N = 0,1,2,3,5 \\
 \mu, \nu = 0,1,2,3\\
 \end{array}
\lbl{4.7}
\ee
where the indices $M,N$ now assume five values only, and
${\cal N}=1/\sqrt{G^{55}}$. The inverse metric is
\be
   G^{MN} = \begin{pmatrix}  {1}/{{\cal N}^2} , & {{\cal N}^\mu}/{{\cal N}^2}& \\
                    {{\cal N}^\nu}/{{\cal N}^2} ,& g^{\mu \nu} + {{\cal N}^\mu {\cal N}^\nu}/{{\cal N}^2}\\
                    \end{pmatrix}
\lbl{4.7a}
\ee                    

The extrinsic curvature is
\be
K_{\mu \nu } \, = \,{\cal D}_\nu  n_\mu  \, 
= \,\frac{1}{{2{\cal N}}}\,\left( {D_\nu  {\cal N}_\mu   + D_\mu  {\cal N}_\nu  
 - \,\frac{{\,\,\partial g_{\mu \nu } }}{{\partial \tau }}} \right)~,
 ~~~~\tau \equiv x^5 .
\lbl{4.8}
\ee
Here ${\cal D}_\nu$ is the 5D covariant derivative, $D_\nu$ the 4D
covariant derivative, and $n_M$ the normal to $M_{1,3}$. 

The 4D metric $g_{\mu \nu}$ depends not only on four spacetime
coordinates $x^\mu$, but also on an extra parameter $\tau$.

Introducing
\be
p^{\mu \nu } \, = \kappa \,\sqrt { - g} \,(Kg^{\mu \nu }  - K^{\mu \nu } ) ,
\lbl{4.9}
\ee
where $\kappa = 1/(16 \pi {\cal G})$,
$g \equiv \det \,g_{\mu \nu }$, and $K \equiv g^{\mu \nu} K_{\mu \nu}$,
we can write the 5D action in the `phase space' form:
\be
I_G [g_{\mu \nu } ,p^{\mu \nu } ,{\cal N},{\cal N}^\mu  ] 
= \int {d\tau \,d^4 x\,
\left[ {p^{\mu \nu } \,\dot g_{\mu \nu } \, 
- \,{\cal N} {\cal H} (g_{\mu \nu } ,p^{\mu \nu } )\, 
- \,{\cal N}_\mu  {\cal H}^\mu  (g_{\mu \nu } ,p^{\mu \nu } )} \right]} ,
\lbl{4.10}
\ee
where
\be
 {\cal H} = \,2\,\kappa \, \sqrt { - g} \,{\cal N}^2 \,G^{55} \, 
 = \kappa \,\sqrt { - g} \,(R^{(4)}  + K^2  - K^{\mu \nu } K_{\mu \nu } )
\lbl{4.11a}
\ee
\be
 {\cal H}^\mu = \,2\,\kappa \, \sqrt { - g} \,{\cal N}\,G^{5\mu } \, 
 = \,2\,D_\nu  p^{\mu \nu } 
\lbl{4.11b}
\ee
The terms with extrinsic curvature in eq.\,(\ref{4.11a}) can be expressed
in terms of $p^{\mu \nu}$:
\be
K^2  - K^{\mu \nu } K_{\mu \nu } \, 
= \,\frac{1}{\kappa^2 (- g) } \,\left( {\frac{{p^2 }}{{D - 1}}\,\, 
- \,p^{\mu \nu } p_{\mu \nu } } \right) ,
\lbl{4.12}
\ee
where $D\, = \,g_{\mu \nu } g^{\mu \nu }  = 4$ and
$p \equiv g_{\mu \nu } \,p^{\mu \nu } = \,\sqrt { - g} \,(D - 1)K$.

Here $p^{\mu \nu}$ are  the  canonical momenta conjugated
to the 4D metric $g_{\mu \nu}$,
whilst ${\cal N}$ and ${\cal N}_\mu$ are Lagrange multipliers for the constraints
\bear
 &&{\cal H}\, = 0 , \lbl{4.13}\\
 &&{\cal H}^\mu   = 0 . \lbl{4.14}
\ear

Upon quantization, $g_{\mu \nu}$ and $p^{\mu \nu}$ become operators
that can be represented as
\be
g_{\mu \nu } \, \to \,g_{\mu \nu } \,,\,\,\,\,
\,\,\,p_{\mu \nu } \, \to \, - \,i\,\frac{\delta }{{\delta g_{\mu \nu } }}
\lbl{4.15}
\ee
More precisely, momentum operator has to satisfy the condition of
Hermiticity, therefore the above definition is not quite correct
in curved spaces, and has to be suitably modified. There also exists the
factor ordering ambiguity that has to be adequately delt with. We are
not interested here into such issues, therefore the expressions with
$-i \delta/\delta g_{\mu \nu}$ have symbolic meaning only.

The `Hamiltonian' constraint,   ${\cal H} \approx 0$, becomes the
Wheeler--DeWitt equation:
\be
   \left [ - \frac{1}{2 \kappa \sqrt{-g}}
   \left (  \frac{{g_{\mu \nu \,} g_{\alpha \beta } }}{{D - 1}}\,\,
  - \,\,g_{\mu \,\alpha } \,g_{\nu \beta } \right )
   \frac{\delta ^2}
     {\delta g_{\mu \nu } \delta g_{\alpha \beta }} 
    + \kappa \, \sqrt{-g} R^{(4)} \right ]
   \Psi [g_{\mu \nu } ]\, = \,0~,~~~~D=4 .
\lbl{4.16}
\ee   

Now the wave function(al) depends on 4-geometry, represented by a
spacetime metric $g_{\mu \nu } \,(x^\mu)$.
In this theory we have no problem of spacetime.
We also have no problem of time, if by `time' we understand the coordinate time
$t\, \equiv \,x^0$.

However, the evolution parameter $\tau$ has disappeared from the
quantized theory. There is no $\tau$ in the wave functional equation
(\ref{4.16}). Now we have the problem of $\tau$.
One possibility is to take the position that this is not a problem.
It is important that we do not have the problem of $t\equiv x^0$           ,         
whereas missing   $\tau$   is not a problem at all.

Another possibility is to bring $\tau$ into the game by considering
matter degrees of freedom. In our approach the latter degrees of
freedom are described by coordinates of Clifford space, one of them being
interpreted as $\tau$.
To describe matter configurations, we have to consider also the matter
part of the action.

As a model we consider the action (\ref{4.1}) in which $R^{(6)}$ is
replaced with $R^{(5)}$, and $\dd^6 x$ with $\dd^5 x$, the indices being
now $M,N = 0,1,2,3,5$. The gravitational part we then replace by the
equivalent phase space action (\ref{4.10}).
The matter part of the action we also replace by the phase space form:
\be
I_{\rm{m}} \, = \,\,\int d \sigma \left( {P_M \dot X^M \, 
- \,\frac{\alpha }{2}(G_{MN} P^M P^N \, - \,\,M_p^2 )} \right)
\lbl{4.17}
\ee
Splitting the metric according to (\ref{4.7}), we have
\be
I_{\rm{m}} \, = \,\,\int d \sigma \left( {P_M \dot X^M \, 
- \,\frac{\alpha }{2}\,\left[ {g_{\mu \nu } (P^\mu   
+ {\cal N}^\mu  P^5 )\,(P^\nu   + {\cal N}^\nu  P^5 ) + \,{\cal N}^2 P^5 P^5  
- \,M_p^2 } \right]} \right)
\lbl{4.18}
\ee
To cast the matter part into a form comparable to the gravitational
part of the action, we insert the integration over
$\delta^5 (x - X(\sigma ))d^5 x$, which gives identity. In both parts
of the action, $I_m$ and $I_G$, now stands the integration over
$\dd^5 x$. Recall that we identified $x^5 \equiv \tau$.

Varying the total action
\be
    I = I_G + I_m
\lbl{4.19}
\ee
with respect to $\alpha$,  ${\cal N}$ and ${\cal N}^\mu$, we obtain the constraints
\bear
    &&\delta \alpha ~: ~~~~~{g_{\mu \nu } (P^\mu   
+ {\cal N}^\mu  P^5 )\,(P^\nu   + {\cal N}^\nu  P^5 ) + \,{\cal N}^2 P^5 P^5
- \,M_p^2 } = 0, \lbl{4.19a} \\
  &&\delta {\cal N}~:~~~~~ {\cal H} + 
   \int \dd \sigma \alpha {\cal N}\delta^5 (x - X(\sigma ))P^5 P^5 \, 
  = 0, 
\lbl{4.20}\\
&&\delta {\cal N}^\mu:~~~~
{\cal H}_\mu  -  
\int \dd \sigma \alpha \,\delta^5 (x - X(\sigma ))\,g_{\mu \nu }
(P^\nu   + {\cal N}^\nu  P^5 ) P^5 = 0.
\lbl{4.21}
\ear
where ${\cal H}$ and ${\cal H}^\mu$ are given in 
eqs.\,(\ref{4.11a}),(\ref{4.11b}), and $\kappa \equiv 16 \pi {\cal G}$.
We can write ${\cal H}$ compactly as 
\be
   {\cal H} = \frac{1}{\kappa}\,
   {\cal G}_{\mu \nu \,\alpha \beta }\, p^{\mu \nu } p^{\alpha \beta } 
              + \kappa \, \sqrt{-g} R^{(4)} ,
\lbl{4.22}
\ee
with the metric
\be
   {\cal G}_{\mu \nu \,\alpha \beta } = \frac{1}{2 \sqrt{-g}} \left [
   \frac{{g_{\mu \nu \,} g_{\alpha \beta } }}{{D - 1}}\,\,
  - \, \frac{1}{2} (g_{\mu \,\alpha } \,g_{\nu \beta } +
  g_{\mu \beta } \,g_{\nu \alpha } ) \right ] ~,~~~~~~D=4.
\lbl{4.24}
\ee

In a quantized theory, the constraints (\ref{4.19a})--(\ref{4.21})
become operator equations acting on a state vector. The constraint
(\ref{4.19a}) can be put straightforwardly into its quantum
version by replacing $P_\mu \rightarrow \hat P_\mu = -i \p_\mu$,
$P_5 \rightarrow \hat P_5 = -i \p_5$. The latter definition of
momentum operator holds in flat space only. In curved space we have
to take a modified definition. For instance, a possible definition
\ci{DeWittHermit} that renders $P_M$ hermitian, and also resolves
the factor ordering ambiguity, is
$\hat P_M = - i[\p_M + (-G)^{-1/4} \p_M (-G)^{-1/4}]$. An alternative
procedure was proposed in Ref.\,\ci{PavsicOrderAmbig}.

So we have
\be
\left[ {g_{\mu \nu } (\hat P^\mu   
+ {\cal N}^\mu  \hat P^5 )\,(\hat P^\nu   + {\cal N}^\nu  \hat P^5 ) 
+ \,{\cal N}^2 \hat P^5 \hat P^5  
- \,M_p^2 } \right] \Psi  = 0 .
\lbl{4.25}
\ee

But the constraints (\ref{4.20}),(\ref{4.21}), because of the
$\delta$-distribution, are not practical for a direct translation into
their corresponding quantum equivalents. Usually, for a quantum description
of gravity in the presence of matter, one does not take the matter action
in the form (\ref{4.18}). Instead, one takes for $I_m$ an action
for, e.g., a scalar or spinor field, and then attempts to quantize the total
action (\ref{4.19}) following the established procedure of quantum
field theory. Here I would like to point out that one can nevertheless
start from the action (\ref{4.18}) and use all the constraints
(\ref{4.19a})--(\ref{4.21}).

Let us consider the Fourier transform of the constraint (\ref{4.20}),
the zero mode being given by the integral
\be
    \int \dd^5 x \,{\cal H} = - \int \alpha \dd \sigma  \, {\cal N} (P^5)^2
\lbl{4.26}
\ee
Writing $\dd^5 x = \dd^4 x \, \dd x^5$ and introducing
$H=\int \dd^4 x {\cal H}$, we have
\be
    \int \dd x^5 \, H = - \int \alpha \dd \sigma \, {\cal N} (P^5)^2,
\lbl{4.27}
\ee
or
\be
     \dd x^5 \,H = - \, \alpha \dd \sigma \, {\cal N} (P^5)^2,
\lbl{4.28}
\ee
from which it follows
\be
  \frac{1}{\alpha} \frac{\dd X^5}{\dd \sigma}\, H = - {\cal N} (P^5)^2.
\lbl{4.29}
\ee
Here we have replaced the coordinate $x^5$, denoting a point in the $5D$
manifold, with the coordinate $X^5$, denoting a point on the worldline.
Using the equation of motion (resulting from varying the action (\ref{4.17})
with respect to $P^M$),
\be
      P^M = \frac{\dot X^M}{\alpha},
\lbl{4.29a}
\ee
where $\dot X^M \equiv \dd X^M/\dd \sigma$,
we find that $P^5 = {\dot X^5}/{\alpha}$. Using the latter expression
in eq.\,(\ref{4.29}), we obtain
\be
    H = - {\cal N} P^5 .
\lbl{4.30}
\ee
Similarly, from the constraint (\ref{4.21}) we obtain
\be
    H_\mu = g_{\mu \nu} (P^\nu + {\cal N}^\nu P^5),
\lbl{4.31}
\ee
where $H^\mu = \int \dd^4 x \, {\cal H}^\mu$. Let us now use the relations
$P^M = G^{MN} P_N$ and $P_M = G_{MN} P^N$ with the metrics
(\ref{4.7}),(\ref{4.7a}), and rewrite eqs.\,(\ref{4.30}),(\ref{4.31}) into
the form with covariant components of momenta $P_\mu,~P_5$:
\be
   H = - \frac{1}{{\cal N}} (P_5 + {\cal N}^\mu P_\mu), \lbl{4.32}
\ee
\be   H_\mu = P_\mu . ~~~~~~~~~~~~~~~~~ \lbl{4.33}
\ee
The above result is nothing but a manifestation of the fact that
the integration of a stress-energy tensor over a certain hypersurface
gives momentum. Here momentum is $P_M = (P_\mu, P_5)$.
Using (\ref{4.11b}), eq.\,(\ref{4.33})
can be rewritten as
\be
   2 \int_\Omega \dd^4 x \, \DD_\nu p^{\mu \nu} 
   = 2 \int_B \dd \Sigma_\nu p^{\mu \nu} = P^\mu~,~~~~~~
     P^\mu \equiv g^{\mu \nu} P_\nu,
\lbl{4.33a}
\ee
where $B$ is the boundary of a region $\Omega$ in the 4-space, and
$\dd \Sigma_\nu$ is an element of the boundary surface. The relation
(\ref{4.33a}) is analogous to the Gauss law in electrodynamics. Bear in mind
that the momentum $P^M$ points along a worldline $X^M (\sigma),~M=(\mu,5)$,
which intersects 4D spacetime in one point. Therefore, the integral in
eq.\,(\ref{4.33a}) is different from zero only when the 3-surface $B$
embraces the intersection point.

For the
Lagrange multipliers we can choose ${\cal N}=1$ and ${\cal N}^\mu =0$, which
simplifies  eqs.\,(\ref{4.32}) and (\ref{4.19a}) into
\be
H = - P_5 ,
\lbl{4.32a}
\ee
\be
  g^{\mu \nu} P_\mu P_\nu +P_5 P_5 - M_p^2 = 0 .
\lbl{4.19b}
\ee    

It is now straightforward to consider the quantum versions of the
constraints (\ref{4.32a}),(\ref{4.33}) together with the constraint
(\ref{4.19b}). We have\footnote{See the texts after Eqs.\,(\ref{4.15})
and (\ref{4.24}).
We chose the factor ordering 
in order to achieve covariance in the space comprised of $X^\mu$.
Therefore, in Eq.\,(\ref{4.34}) we have the covariant derivative
$\DD/\DD X^\mu$.  In an analogous way
should be interpreted Eq.\,(\ref{4.35}).  }
\be
    \left ( g^{\mu \nu} \frac{\DD^2}{\DD X^\mu \DD X^\nu}
     + \frac{\p^2}{\p \tau^2} + M_p^2\right ) \Psi = 0
\lbl{4.34}
\ee
\be
    \int \dd^4 x \left ( - \frac{1}{\kappa} G_{\mu \nu \, \alpha \beta}
   \frac{\delta ^2}
     {\delta g_{\mu \nu } \delta g_{\alpha \beta }} + \kappa \sqrt{-g} R^{(4)} \right )
   \Psi  =  i \, \frac{\p}{\p \tau} \Psi ~ ,~~~~~~\tau \equiv X^5 .
\lbl{4.35}
\ee
\be
    \int \dd^4 x \, \DD_\nu
   \left( -i \frac{\delta}{\delta g_{\mu \nu}} \right ) \Psi =
   - i \frac{\p}{\p X^\mu} \Psi
\lbl{4.36}
\ee   
The latter equations impose the operator constraints on a quantum
state that is
represented by $\Psi[\tau, X^\mu, g_{\mu \nu}(x^\mu)]$ which depends
on the particle's coordinates $X^\mu$, the fifth coordinate
$X^5 \equiv \tau$, and the spacetime metric $g_{\mu \nu}(x^\mu)$.
In other words, $\Psi$ is a function of $\tau, ~ X^\mu$, and
a functional of $g_{\mu \nu}(x^\mu)$. Eq.\,(\ref{4.35}) is just like the
Schr\"odinger equation, with $\tau$ as evolution parameter.
 Therefore, the ``problem of $\tau$''
does not exist in this quantum model for a point particle coupled
to a gravitational field. Had we performed a split
from six to four dimensions (and not from five to four as we did in this
section), then in eq.\,(\ref{4.34}), instead of $\p_\tau^{\,2}$,  we would have 
$\p_\lambda \p_\tau \sim \Lambda \p_\tau$ (see sec.\,2), so that
eq.\,(\ref{4.34}) would become the Stueckelberg equation.

The system (\ref{4.34})--(\ref{4.36}) describes at once a Klein-Gordon
wave function for a relativistic particle, and the wave functional
for a gravitational field. It is only an incomplete description of
the  physical system. A complete description would require to take
into account the infinite set of constraints due to all Fourier modes of the
the constraints (\ref{4.19a})--(\ref{4.21}). 

\section{Discussion and conclusion}

We have shown how the Stueckelberg equation for a relativistic
point particle comes from a 6-dimensional space, $M_{2,4}$, 
with signature (2,4), that is $(+ - - - -+)$.
Two extra dimensions, one time like and one space like, are necessary,
because then in the equation we obtain
the first derivative of the wave function
with respect to a Lorentz, SO(1,3), invariant parameter $\tau$ which is
identified with the fifth coordinate $X^5$.

An argument in favor of such 6D space comes from the works on the two time (2T)
physics\,\ci{Bars} that
is invariant under local Sp(2) transformations between coordinates
and momenta. In such theory there are three Lagrange multipliers associated
with three constraints, which cannot be satisfied in 4D spacetime
$M_{1,3}$. They can be satisfied in 6D space $M_{2,4}$, or in a suitable
higher dimensional space.
Since the theory by Bars et al.\,\ci{Bars} is based on very strong
foundations, we can conclude that the 6D space is a reasonable
subsitute for 4D spacetime. It enables to formulate
the 2T physics on the one hand, and the Stueckelberg theory on the other hand,
but not both at once.
A relationship between the two theories has yet to be
explored. A clue is to consider a higher than six dimensional space and to
impose the Sp(2) constraints on the variables entering the phase space
action, and thus obtain a generalization of the 2T physics.
The phase space action (\ref{2.9}) in six dimensions---that the Stueckelberg
theory in embedded in---is a particular, gauge fixed, case
of the Sp(2) invariant action in higher than six dimensions. 
According to such view, the local Sp(2) invariance holds in a higher dimensional
space, whereas in the 6-dimensional
subspace $M_{2,4}$,  it is broken. But, in $M_{2,4}$ one might expect the
problem with ghosts due to the extra time like dimension.
Concerning ghosts, it was shown
in Refs.\,\ci{PavsicSaasFee,PavsicPseudoHarm,PavsicBook}
that they do not necesarily occur in spaces with
time like dimensions, if one defines vacuum in an alternative way,
as proposed by Jackiw et al.\,\ci{Jackiw}.

There exists another direction of research, which is based on the concept of
configuration space, i.e., the space of possible matter configurations.
An example of such space is the 16D space of oriented $r$-volumes,
associated with extended objects, e.g., branes. We call it Clifford space, $C$,
because it is a manifold whose tangent space at any point is a
Clifford algebra $Cl(1,3)$. If we define the metric according to
eq.\,(\ref{3.6}), then the signature of $C$ is $(8,8)$. A subspace of
$C$ is $M_{2,4}$. Therefore, if we adopt the concept of Clifford
space, $C$, we do not need to postulate extra dimensions of spacetime,
in order to have the 6D space formulation of the Stueckelberg theory, or
of the 2T physics. Four dimensions of $C$ can be identified with the
four dimensions of spacetime, whilst the remaining 12 dimensions of $C$
are associated with the intrinsic configurations of matter living
in the 4-dimensional spacetime.

We have considered the general relativity in Clifford space, more precisely
in the 6D subspace with signature (2,4). The action contains the
Einstein-Hilbert term which is a functional of the metric only,
and a matter term, which is a functional of matter degrees
of freedom coupled to the metric. As a model we have considered a point
like source. We have performed the ADM decomposition of a 5D subspace into
the spacetime $M_{1,3}$ and a part due to the 5th dimensions, $x^5$.
The action gives the mass shell constraint in 5-dimensions,
and the constraints that generalize
the Hamiltonian and momentum constraints of the canonical gravity, with
the extra terms due to the presence of the point particle source.
After quantization those constraints become the operator constraints
acting on a state that can be represented as a functional of the
spacetime metric $g_{\mu \nu},~\mu, \nu=0,1,2,3$, a function of the particle
coordinates $X^\mu$, and the fifth coordinates, $X^5 \equiv \tau$,
which has the role of the Stueckelberg evolution parameter. In the Stueckelberg
theory the `true' time is the Lorentz, SO(1,3), invariant evolution
parameter $\tau$, and not the coordinate $x^0 \equiv t$. Since such
parameter occurs in the wave function(al) for the gravitational field,
we conclude that there is no `problem of time' in this theory.

\vs{3mm}

\centerline{\bf Acknowledgment}

\vs{1mm}

This research was supported by the Ministry of High Education,
Science and Technology of Slovenia.

\end{document}